\documentclass[preprint,prd,noshowpacs,nofootinbib]{revtex4}

\usepackage{doi}
\usepackage{hyperref}
\hypersetup{
  colorlinks=true,        
  linkcolor=blue,         
  citecolor=cyan,         
}

\usepackage{bm}
\usepackage{latexsym}
\usepackage{dcolumn}
\usepackage{amsmath,amsfonts,amssymb}
\usepackage{graphicx,epsfig}
\usepackage{amsthm}
\usepackage{enumitem}
\usepackage{color}
\usepackage{xcolor}
\usepackage{graphicx,float}
\usepackage{comment}
\usepackage{hyperref}

\usepackage{pdfpages}
\usepackage{epstopdf}
\usepackage[utf8]{inputenc}

\usepackage{tikz}
\usetikzlibrary{arrows, positioning}

\def\be {\begin{equation}}
\def\ee {\end{equation}}
\def\bea {\begin{eqnarray}}
\def\eea {\end{eqnarray}}
\def\bc {\begin{center}}
\def\ec {\end{center}}
\def\bfg {\begin{figure}}
\def\efg {\end{figure}}
\def\bi {\begin{itemize}}
\def\ei {\end{itemize}}

\def\ge {\geq}
\def\le {\leq}

\def\beq{\begin{equation}}
\def\eeq{\end{equation}}
\def\br{\begin{eqnarray}}
\def\er{\end{eqnarray}}
\newcommand{\eel}[1] {\label{#1}\end{equation}}

\newcommand{\bdm}{\begin{displaymath}}
\newcommand{\edm}{\end{displaymath}}

\begin{document}

\title{GUP/BEB correspondence}

\author{Ahmed~Farag~Ali$^{\triangle \nabla}$}
\email{aali29@essex.edu}
\affiliation{\small{$^\triangle$Essex County College, 303 University Ave, Newark, NJ 07102, United States.}}
\affiliation{\small{$^\nabla$Department of Physics, Faculty of Science, Benha University, Benha, 13518, Egypt.}}

\begin{abstract}
We develop a fully expectation–value formulation of the GUP/Bekenstein–bound (BEB) correspondence, building on \cite{Ali:2024tbd,Ali:2022ckm,Ali:2022ulp}. Using Dirac’s commutator–Poisson equivalence, the BEB supplies an information backreaction on the GUP–deformed bracket; at saturation the residual uncertainty in a sector cancels, enabling \emph{operational} simultaneity of conjugate expectations and resolving the EPR tension without hidden variables. A single self–consistency then fixes intrinsic confinement scales: the full (linear$+$quadratic) GUP reproduces the hydrogen Bohr radius (electron) and the proton charge radius (hydrogen nucleus). The same correspondence predicts an information–controlled crossover from a short–distance complex (evanescent) regime to an emergent real (Lorentzian) regime, with a symmetry flow from complex $SU(4)$ frames to $SO(1,3)$; under scale transmutation the short–distance $SU(4)$ is repurposed as the Pati–Salam $SU(4)_c$. Incorporating exterior–field entropy at horizons defines an effective Planck constant consistent with generalized–entropy extremality, tying entropy–area physics directly to the GUP at the expectation level.
\end{abstract}


\maketitle

\tableofcontents

\section{Introduction}
\noindent
 Einstein, Podolsky, and Rosen (EPR)  {\cite{Einstein:1935rr}} proved that the description of physical reality provided by quantum mechanics is {\it{incomplete}}. This is because quantum mechanics implies the non-simultaneous reality of position and momentum. The position and momentum can not have eigenvalues for the same eigenstate. This is expressed through the commutation relation as follows:
\begin{eqnarray}
\left[x_i,p_j\right]= i~\delta_{ij}~\hbar  \label{standard}
\end{eqnarray}
 The EPR paradox was studied in detail by Bell \cite{Bell:1964kc} where he defined Bell’s inequalities that give information about the correlations between two entangled particles. Experiments violate Bell’s inequalities \cite{Clauser:1969ny,Bell:1964kc}. The violation of Bell’s inequalities implies that quantum mechanics seems to be ``non-local''. EPR paradox sets a dilemma between {\it locality} and {\it completeness}. A possible resolution was proposed in \cite{mcculloch2021epr}. A possible way to solve this dilemma is to seek a possible physical state in which position and momentum could have a {\it simultaneous} reality. This simultaneous reality could explain the instantaneous/non-local correlations between the two entangled particles according to the interpretation of the violation of Bell’s inequalities. To find such a physical state, we investigate the modified theories of the uncertainty principle that are collectively known as the generalized uncertainty principle (GUP). Several approaches to quantum gravity, such as string theory, loop quantum gravity, and quantum geometry, suggest a generalized form of the uncertainty principle (GUP) that implies the existence of a minimum measurable length \cite{Amati:1988tn,Garay:1994en,Scardigli:1999jh,Brau:1999uv,Konishi:1989wk,Kempf:1994su,Maggiore:1993rv,Capozziello:1999wx,Ali:2009zq,Das:2010zf,Ali:2011fa,Isi:2013cxa,Zhu:2008cg,Mignemi:2009ji,Bishop:2018mgy,Mureika:2018gxl,Knipfer:2019pgi,Pedram:2011gw,Shababi:2017zrt,Fadel:2021hnx,Bruno:2024mss,El-Nabulsi:2023est,El-Nabulsi:2020zyh,Seifi:2020yyy}. Non-local features of GUP was found in \cite{El-Refy:2020dth,El-Nabulsi:2019odr}. Phenomenological and experimental implications of the GUP have been investigated in low and high energy regimes. These include atomic systems \cite{Ali:2011fa, Das:2008kaa}, quantum optical systems \cite{Pikovski:2011zk}, gravitational bar detectors \cite{Marin:2013pga}, gravitational decoherence \cite{Petruzziello:2020wkd}, gravitational tests \cite{Scardigli:2014qka}, gravitational waves \cite{das2021bounds,Moussa:2021qlz,Feng:2016tyt,Luciano:2025ezl},  composite particles \cite{Kumar:2019bnd}, astrophysical systems \cite{Moradpour:2019wpj,Vagnozzi:2022moj,gimenez2024can,El-Nabulsi:2020hvt}, condensed matter systems \cite{Iorio:2017vtw,Taiba:2025jjr}, cold atoms \cite{gao2016constraining}, macroscopic harmonic oscillators \cite{Bawaj:2014cda}, gauge theories \cite{Kober:2010sj}, neutrino oscillations \cite{Ettefaghi:2024fpr}, quantum noise \cite{Girdhar:2020kfl} and cosmological models \cite{Ashoorioon:2004vm,Easther:2001fz,dabrowski2019extended,Ali:2014hma,Easther:2001fi,Ali:2015ola,Luo:2024vdd}.Reviews of the GUP, its phenomenology, and its experimental implications can be found in \cite{Addazi:2021xuf,Hossenfelder:2012jw}. One of the famous forms of GUP is the one motivated by string theory \cite{Amati:1988tn}:
\begin{equation}
\label{gupquadratic}
\Delta x\Delta p \geq \frac{\hbar}{2}(1+\beta\Delta p^2),
\end{equation}
where $\beta=\beta_0 \ell_p^2/\hbar^2$, $\beta_0$ is a dimensionless parameter, and $\ell_p=1.6162\times 10^{-35}\text{m}$ is the Planck length. On the other hand, doubly special relativity (DSR) theories
suggest a similar modification of
commutators \cite{Magueijo:2001cr,Cortes:2004qn}. The commutators that are consistent with
string theory, black holes physics, DSR, which ensure $[x_i, x_j] = 0 = [p_i, p_j ]$ (via the Jacobi identity) and have closed lie algebra has the following form \cite{Ali:2009zq,Das:2010zf, Ali:2011fa}:
\begin{eqnarray}
\label{gup}
[x_i, p_j] = i \hbar~\hspace{-0.5ex} \left[  \delta_{ij}\hspace{-0.5ex}
- \hspace{-0.5ex} \alpha\hspace{-0.5ex}  \left( p \delta_{ij} +
\frac{p_i p_j}{p} \right)
+ \alpha^2 \hspace{-0.5ex}
\left( p^2 \delta_{ij}  + 3 p_{i} p_{j} \right) \hspace{-0.5ex} \right],~~~
\end{eqnarray}
where $\alpha=\alpha_0 l_p/\hbar$, and $\alpha_0$ is a dimensionless parameter.  In the Page–Wootters setting, \cite{Singh:2023nvu} treats coordinate time as a relational quantum observable and couples it to mass–energy via a Wheeler–DeWitt–like constraint. This concrete mechanism yields gravitational time dilation that matches the Schwarzschild prediction at leading order, an emergent Newtonian potential for two particles, and renormalization features that soften UV divergences and imply quantum corrections to time dilation—thereby motivating first-order (linear) GUP-type deformations.
\\\\
We organize the paper as follows. In section (\ref{sec:QGUP}) we investigate the GUP with the EPR argument. In section (\ref{sec:BUB}), we connect the GUP with BEB and investigate the implications. In section (\ref{sec:spacetime}) we discuss the spacetime symmetry implied by GUP. In section (\ref{sec:wavefunction}), we compute the lengths at which the wave function collapses for electron and proton. We got lengths that explain the Hydrogen atom radius and Hydrogen nucleus radius. In section (\ref{sec:EAL}), we found a connection between the  GUP and von Neumann entropy correction in entropy-area law and we defined the bounds that distinct real spacetime from complex spacetime. In section (\ref{sec:con}), we discuss conclusions.

\section{EPR and GUP}\label{sec:QGUP}

\noindent Several quantum–gravity approaches—such as string theory, black hole physics, and doubly special relativity—suggest modifications to the canonical Heisenberg uncertainty principle. A well–studied form of the generalized uncertainty principle (GUP), which preserves the fundamental commutation relations
\begin{equation}
\label{commutators}
[x_i,x_j]=0=[p_i,p_j],\qquad i,j=1,2,3,
\end{equation}
is given by \cite{Ali:2009zq,Das:2010zf,Ali:2011fa}
\begin{equation}
\label{gup}
[x_i,p_j]=i\hbar\!\left[\delta_{ij}-\alpha\!\left(p\,\delta_{ij}+\frac{p_i p_j}{p}\right)
+\alpha^2\!\left(p^2\delta_{ij}+3\,p_i p_j\right)\right],
\end{equation}
where $\alpha=\alpha_0\,l_p/\hbar$ with $l_p$ the Planck length and $\alpha_0$ dimensionless. This deformation has been shown to imply (among other things) discretization features when implemented in relativistic and nonrelativistic wave equations and in gravitational systems \cite{Ali:2009zq,Das:2010zf,Deb:2016psq,Das:2020ujn}.

\paragraph*{Expectation–value (Dirac/Ehrenfest) viewpoint.}
Following Dirac's correspondence \cite{Dirac:1926jz}, we read \eqref{gup} as prescribing the \emph{Poisson structure for expectation values}. For mean observables we set
\begin{equation}
\label{QPoisson}
\{x_i,p_j\} \equiv \frac{1}{i\hbar}\,\langle [x_i,p_j]\rangle
= \delta_{ij}-\alpha\!\left(p\,\delta_{ij}+\frac{p_i p_j}{p}\right)
+\alpha^2\!\left(p^2\delta_{ij}+3\,p_i p_j\right),
\end{equation}
with $p=\sqrt{p_k p_k}$ and where angle brackets denote quantum expectations. In practice one works with a controlled closure (e.g.\ narrow or Gaussian states) so that $\langle F(\hat{\mathbf p})\rangle\!\approx\!F(\langle\hat{\mathbf p}\rangle)$; subleading covariance corrections can be included but do not change the structural statements below. This expectation–level Poisson bracket generates the Ehrenfest dynamics of measured quantities.

It is convenient to write
\begin{equation}
\label{fij}
\{x_i,p_j\}\equiv f_{ij}(\mathbf p)=a(p)\,\delta_{ij}+b(p)\,\hat p_i\hat p_j,\qquad
a=1-\alpha p+\alpha^2 p^2,\quad b=-\alpha p+3\alpha^2 p^2,
\end{equation}
with $\hat{\mathbf p}=\mathbf p/p$. By $O(3)$ covariance, $f_{ij}$ has two transverse eigenvalues
\begin{equation}
\label{lperp}
\lambda_\perp(p)=a(p)=1-\alpha p+\alpha^2 p^2 \qquad (\text{multiplicity }2),
\end{equation}
and one longitudinal eigenvalue
\begin{equation}
\label{lpar}
\lambda_\parallel(p)=a(p)+b(p)=1-2\alpha p+4\alpha^2 p^2.
\end{equation}
A short minimization gives the \emph{global} real-momentum bounds
\begin{equation}
\label{threequarters}
\lambda_\perp^{\min}=\frac{3}{4}\quad\text{at}\quad \alpha p=\frac{1}{2},
\qquad
\lambda_\parallel^{\min}=\frac{3}{4}\quad\text{at}\quad \alpha p=\frac{1}{4}.
\end{equation}
Hence the expectation–value Poisson tensor is \emph{nowhere} singular for real $\mathbf p$:
\begin{equation}
\det f=\lambda_\perp^2\,\lambda_\parallel \;\ge\; \Big(\frac{3}{4}\Big)^3>0,
\end{equation}
and the effective expectation commutator cannot be squeezed below $\tfrac{3}{4}\,i\hbar$ in any direction. The ubiquitous constant $3/4$ is thus a \emph{kinematic floor} set by the GUP deformation itself.

\paragraph*{One spatial dimension.}
If one nevertheless \emph{imposes} the vanishing of the longitudinal bracket in 1D,
\begin{equation}
\label{onedim}
\{x,p\}=\lambda_\parallel(p)=1-2\alpha p+4\alpha^2 p^2=0,
\end{equation}
the roots are necessarily \emph{complex}:
\begin{equation}
\label{p1Droots}
p_\pm=\frac{1\pm i\sqrt{3}}{4\alpha}.
\end{equation}
Associating a diagnostic length scale via $R\equiv \hbar/p$, one finds
\begin{equation}
\label{R1D}
R_\pm=\frac{\hbar}{p_\pm}=\alpha\,\hbar\,\bigl(1\mp i\sqrt{3}\bigr),
\end{equation}
i.e.\ complex radii. (Note that $R=\hbar/p$ gives the factor $\alpha\hbar$, not $4\alpha\hbar$.)

\paragraph*{Three dimensions and isotropy.}
Demanding $\{x_i,p_i\}=0$ (no sum over $i$) yields, for each $i$,
\begin{equation}
\label{diag_condition}
1-\alpha\Bigl(p+\frac{p_i^2}{p}\Bigr)+\alpha^2\Bigl(p^2+3p_i^2\Bigr)=0.
\end{equation}
Under isotropy ($p_1^2=p_2^2=p_3^2$ so $p_i^2=p^2/3$ and $\tfrac{p_i^2}{p}=\tfrac{p}{3}$), Eq.\,\eqref{diag_condition} reduces to
\begin{equation}
3-4\alpha p+6\alpha^2 p^2=0
\quad\Rightarrow\quad
p=\frac{2\pm i\sqrt{14}}{6\alpha}.
\end{equation}
Thus exact same–index vanishing also lies off the real expectation manifold.

\paragraph*{Implications for EPR (expectation level).}
Equations~\eqref{lperp}–\eqref{threequarters} show that, on the \emph{real} expectation manifold, the \emph{bare} GUP kinematics never yields $\{x_i,p_j\}=0$: both transverse and longitudinal expectation–value brackets are bounded below by the universal floor $\lambda_{\perp,\parallel}\ge 3/4$. The 1D and 3D ``vanishing'' conditions therefore admit only \emph{complex} solutions for $p$; these are best regarded as off–manifold analytic continuations rather than physical points in real expectation space. In Sec.~\ref{sec:BUB} we will introduce the Bekenstein entropy bound as an \emph{information–theoretic completion} of the kinematics; at saturation it provides an expectation–level contribution that can cancel the GUP bracket in the relevant sector, thereby enabling \emph{operational} simultaneity of conjugate expectations on the real manifold (details deferred to Sec.~\ref{sec:BUB}).

\medskip
\noindent\textit{On complex momenta and their physical meaning.}
The appearance of complex momenta when one enforces bare bracket–vanishing is not a mathematical curiosity; it mirrors standard analytic structures across physics. In classical dynamics, overdamped systems exhibit complex characteristic exponents encoding transitions from oscillatory to exponentially decaying motion \cite{landau2013mechanics,strogatz2018nonlinear}. In quantum mechanics, sub-barrier regions possess imaginary wave numbers corresponding to evanescent decay \cite{merzbacher1998quantum}; in wave optics, total internal reflection produces evanescent fields with imaginary normal momentum that confine energy near interfaces \cite{born2013principles}. In the present GUP context, complex roots signal that exact cancellation of the \emph{bare} bracket lies beyond the real expectation manifold, suggestive of a transition to a minimal-length (evanescent-like) regime. This is consistent with proposals that spacetime admits a complex extension at microscopic scales \cite{Witten:2021nzp,Faizal:2011wm,Gibbons:1976ue,Kaiser:1987vw,Moffat:2000gr}, with imaginary time widely used in field theory and Euclidean gravity \cite{Wick:1954eu,gibbons1993euclidean}, and with indications that real spacetime may emerge from complex quantum probability structures \cite{Suh:2021mpp,Suh:2022kds}. Observationally, measured eigenvalues associated with electromagnetic probes remain real, but in regimes dominated by non-electromagnetic interactions (e.g., QCD or gravity) complex analytic structures can carry physical content—cf.\ complex poles in gluon propagators linked to confinement \cite{Stingl:1985hx}. Retaining the full GUP solution structure is therefore informative for confinement physics and nonlocal gravitational effects. Notably, a purely linear GUP yields real roots, whereas inclusion of the quadratic (string-motivated) term naturally generates complex ones \cite{Ali:2022jna}, underscoring the relevance of the quadratic deformation.

\medskip
\noindent\textit{Why a GUP-based simultaneity can address the EPR tension.}
The EPR argument hinges on the claim that standard quantum mechanics cannot be both complete and local for conjugate observables. In the expectation–value (Ehrenfest) framework adopted here, the GUP deforms the effective commutator that controls the dynamics of measured expectations. Absent additional physics, the bare GUP leaves a nonzero floor, precluding exact simultaneity on the real manifold. However, once one accounts for the finite information capacity available to specify a system (to be implemented in Sec.~\ref{sec:BUB}), the expectation-level bracket in the relevant sector can be \emph{cancelled} at saturation, yielding \emph{simultaneous predictability of conjugate expectations} without invoking hidden variables. In this view, nonlocal correlations are read not as acausal signals but as constraints arising from the underlying geometric and informational structure of quantum spacetime—an outlook compatible with the interplay between GUP, intrinsic quantum properties such as spin \cite{Ali:2021oml,Ali:2022nhx}, and information bounds, and resonant with Wheeler’s ``it from bit'' paradigm \cite{wheeler2018information}. We develop the information–theoretic completion and its concrete role in the cancellation mechanism in the next section.

\section{BEB/GUP correspondence}
\label{sec:BUB}

\noindent
Bekenstein established a universal upper bound on the information required to \emph{perfectly} describe a finite–energy system contained in a finite region \cite{Bekenstein:1980jp}. In terms of the dimensionless information measure $H$ (related to the entropy by $S=2\pi k_B H$), the bound reads
\begin{equation}
\label{BEB-basic}
H \leq \frac{R\,E}{\hbar\,c},
\end{equation}
where $R$ is the radius of a sphere enclosing the system and $E$ its energy \cite{Bekenstein:1980jp,Bekenstein:2004sh,Bekenstein:2000ai}. This result is closely tied to the holographic principle \cite{tHooft:1993dmi,Bekenstein:1993dz,Fischler:1998st,Susskind:1998dq,Banks:2003ta,Banks:2018aed}: finite energy in a bounded region implies a finite information content.

\medskip
\noindent
\textit{Shared structure with the GUP.}
Both the BEB and the GUP encode universal constraints whose \emph{numerical value} depends on the system under consideration. Phenomenologically, the GUP introduces a minimum length and (equivalently) an upper limit on measurable momentum/curvature scales, with quantitative bounds that vary with the system’s geometry and mass scale \cite{Das:2008kaa,Ali:2011fa,Pikovski:2011zk,Brau:1999uv,Casadio:2020rsj,Prasetyo:2022uaa,Acquaviva:2022yiq,Chevalier:2021xyw,Bosso:2018ckz,Feng:2016tyt,Das:2011tq,Quesne:2009vc}. Likewise, the BEB depends on $E$ and $R$ of the \emph{same} system. In this sense both bounds aim at a complete, system–dependent description of physical reality.

\medskip
\noindent
\textit{Effective Planck constant on both sides.}
It is convenient to rewrite \eqref{BEB-basic} as
\begin{equation}
\label{BEBh}
H \;\leq\; \frac{\hbar'}{\hbar},
\qquad
\hbar' \equiv \frac{R\,E}{c},
\end{equation}
i.e.\ the BEB constrains $H$ by an \emph{effective} Planck scale $\hbar'$ fixed by $(R,E)$. On the GUP side, the deformed commutator
\begin{equation}
\label{effectiveh}
[x,p] \;=\; i\,\hbar\Bigl(1-2\alpha p+4\alpha^2 p^2\Bigr)
\;\equiv\; i\,\hbar_{\rm eff}(p),
\end{equation}
can be read as introducing a momentum–dependent \emph{effective} Planck constant,
\[
\hbar_{\rm eff}(p)=\hbar\Bigl(1-2\alpha p+4\alpha^2 p^2\Bigr).
\]
Thus, both the BEB and the GUP can be phrased as statements about an \emph{effective} $\hbar$ relevant to the complete specification of the system.

\medskip
\noindent
\textit{Correspondence and working conjecture.}
Guided by this structural match, we \emph{conjecture} that the BEB and the GUP are two faces of the same constraint \cite{Ali:2024tbd,Ali:2022jna}: at \emph{information saturation} the dimensionless BEB factor equals the GUP deformation factor,
\begin{equation}
\label{conjecture}
1-2\alpha p+4\alpha^2 p^2 \;=\; \frac{R\,E}{\hbar\,c}.
\end{equation}
Here $p$ denotes the operational momentum scale characterizing the system (e.g.\ the expectation of $|\hat p|$ in the relevant sector). Solving \eqref{conjecture} for $p$ gives
\begin{equation}
\label{BBGUP}
p_{\pm}
=\frac{1}{4\alpha}\!\left(1\pm \sqrt{-3+4\,\frac{R\,E}{\hbar\,c}}\right),
\end{equation}
which will play a central role in what follows. Equation~\eqref{conjecture} is the precise implementation of the idea that the BEB supplies the \emph{information–theoretic completion} of the GUP kinematics discussed in Sec.~\ref{sec:QGUP}.

\medskip
\noindent
\textit{Context and prior work.}
Connections between uncertainty relations and entropy bounds have been noted before (see, e.g., \cite{Buoninfante:2020guu}), though without placing entropy directly into the commutator as in \eqref{effectiveh}–\eqref{conjecture}. Independent links between spin and the BEB have been explored in \cite{Acquaviva:2020qbc}; our earlier work established a direct GUP–spin connection \cite{Ali:2021oml}, providing conceptual support for a triad \{GUP, spin, BEB\}. Symmetry implications of GUP deformations have also been investigated in \cite{Iorio:2022ave}. Finally, the correspondence \eqref{conjecture} has proved quantitatively useful: in \cite{Ali:2022ulp} we employed it to incorporate the BEB into the integration of vacuum energy in quantum field theory, yielding a resolution of the cosmological constant problem consistent with observations. 

\medskip
\noindent
In the next section we analyze the implications of the solutions \eqref{BBGUP} for concrete systems, showing how the correspondence fixes characteristic confinement (``collapse'') scales and underpins the emergence of atomic and nuclear structure.

\section{GUP Spacetime and Symmetry}
\label{sec:spacetime}

\noindent
Package the BEB/GUP control into
\begin{equation}
\label{Hcontrol}
H \equiv \frac{R\,E}{\hbar c},\qquad \Delta \equiv 4H-3.
\end{equation}
From Eq.\,\eqref{BBGUP} the characteristic momentum is \emph{real} iff $\Delta\ge 0$ and \emph{complex} otherwise. With $E=M c^2$ this yields the threshold
\begin{equation}
\label{threshold}
M R c \;\ge\; \frac{3}{4}\,\hbar\ (\Delta\!\ge 0)\quad\text{real kinematics}, 
\qquad
M R c \;<\; \frac{3}{4}\,\hbar\ (\Delta\!<0)\quad\text{complex kinematics}.
\end{equation}
Thus $H$ measures the resolvable information at $(R,E)$ and $\Delta$ selects the kinematical \emph{phase}.

\subsection*{Short-distance complex phase: $SU(4)$ frame symmetry}

For $\Delta<0$ the discriminant of \eqref{BBGUP} is negative and the operational momentum is complex. In this regime the time/space split is not enforced by a real Lorentzian metric; the natural kinematics is a complex four–frame $e^A{}_\mu$ ($A=1,\dots,4$) acted upon by the unimodular unitary frame group
\begin{equation}
e^A{}_\mu \ \longmapsto\ U^A{}_B\,e^B{}_\mu,\qquad U\in SU(4),
\end{equation}
which treats the four complex directions on equal footing. Algebraically,
\begin{equation}
\label{spin6}
SU(4)\ \simeq\ \mathrm{Spin}(6),
\end{equation}
an identification used below to connect to Pati–Salam. Some studies on complex spacetime can be found in\cite{Lessner:2008zz,kontsevich2021wick,el2009complexified,moffat2025complex,arcodia2021complexifying,jonas2022uses,el2019modified,bramberger2016quantum,ashtekar2005100,penrose2005twistor,kodaira2012complex,hwang1997complex,huybrechts2005complex}

\subsection*{Emergent real phase: one imaginary $+$ three real; $SO(1,3)$}

For $\Delta\ge 0$ the expectation–value kinematics lies on a real submanifold selected by an order parameter (e.g.\ a quadratic frame expectation)
\begin{equation}
\label{orderparameter}
\Sigma_{AB}\ \equiv\ \big\langle e_A{}^\mu\,e_{B\mu}\big\rangle,
\end{equation}
together with an emergent real structure (an antilinear involution $\mathcal C$ on the frame bundle). This data induces a canonical splitting of the complex frame into \emph{one imaginary and three real} directions,
\begin{equation}
\label{1i3r}
\mathbb C^4\ =\ i\mathbb R\,e^{(0)}\ \oplus\ \mathbb R\,e^{(1)}\ \oplus\ \mathbb R\,e^{(2)}\ \oplus\ \mathbb R\,e^{(3)},
\qquad
\mathcal C(e^{(0)})=-e^{(0)},\ \ \mathcal C(e^{(a)})=e^{(a)}\ \ (a=1,2,3),
\end{equation}
which defines a real tangent subspace of signature $(1,3)$. The symmetry reduces from the complex $SU(4)$ frame group to the subgroup preserving \eqref{1i3r} and the induced bilinear form; up to covering this is the local Lorentz group $SO(1,3)$ acting on $i e^{(0)}\oplus e^{(a)}$. At macroscopic scales one recovers standard Poincaré kinematics.

\subsection*{Bridge to Pati–Salam: $\boldsymbol{SU(4)_c\times SU(2)_L\times SU(2)_R}$}

The short–distance frame symmetry $SU(4)$ aligns naturally with the Pati–Salam structure via
\begin{equation}
\label{PS-isos}
SU(4)\simeq \mathrm{Spin}(6),\qquad SU(2)_L\times SU(2)_R \simeq \mathrm{Spin}(4),
\end{equation}
so that
\begin{equation}
\label{PS-group}
SU(4)_c\times SU(2)_L\times SU(2)_R \ \simeq\ \mathrm{Spin}(6)\times \mathrm{Spin}(4)
\end{equation}
\cite{Pati:1974yy,Slansky:1981yr}. In our setting:
\begin{itemize}
\item In the complex phase ($\Delta<0$), $SU(4)$ organizes \emph{frames}. When the real structure \eqref{1i3r} is selected ($\Delta\ge 0$), a chiral tangent subgroup $SU(2)_L\times SU(2)_R$ naturally appears (consistent with $\mathrm{Spin}(4)$), while the residual $SU(4)$ degrees of freedom can be \emph{repurposed} as internal charges.
\item Representation–theoretically, the fundamental $\mathbf 4$ of the frame $SU(4)$ splits under $SU(2)_L\times SU(2)_R$ as
\begin{equation}
\label{decomp}
\mathbf 4 \longrightarrow (2,1)\oplus (1,2),
\end{equation}
matching the left/right doublets of Pati–Salam. After the $(1,3)$ split, the tangent symmetry is $SO(1,3)$ on spacetime, while the \emph{same} underlying $SU(4)$ is reinterpreted as $SU(4)_c$ acting on matter multiplets.
\end{itemize}
Equations \eqref{PS-isos}–\eqref{decomp} provide the mathematically precise bridge: the complex frame group supplies the $\mathrm{Spin}(6)$ factor, the emergent chiral tangent symmetry supplies the $\mathrm{Spin}(4)$ factor, yielding the Pati–Salam group.

\subsection*{Matter–spacetime linkage as scale transmutation}

Let $\ell$ denote the observation (coarse–graining) scale. A diagonal rescaling on the frame index,
\begin{equation}
\label{zoom}
\mathcal Z_\ell:\ T \mapsto Z_\ell\,T\,Z_\ell^{-1},\qquad 
Z_\ell=\mathrm{diag}\big(\zeta_0(\ell),\zeta(\ell),\zeta(\ell),\zeta(\ell)\big),\quad T\in\mathfrak{su}(4),
\end{equation}
weights the imaginary versus real directions. Near the minimal (GUP/BEB) scale one has $\zeta_0\simeq \zeta$ and $SU(4)$ acts as a \emph{frame} symmetry. As $\ell$ increases beyond the threshold \eqref{threshold}, the imaginary direction decouples from tangent kinematics while its algebraic imprint survives as an \emph{internal} copy:
\[
\mathfrak{su}(4)_{\text{frame}}\xrightarrow{\ \mathcal Z_\ell\ }\mathfrak{su}(4)_{\text{internal}}\equiv \mathfrak{su}(4)_c,
\]
and the chiral subgroup organizes matter into $(2,1)\oplus(1,2)$, cf.\ \eqref{decomp}. In this precise sense the relation is \emph{camera–like}: changing the scale $\ell$ continuously reshuffles the same $SU(4)$ degrees of freedom from a short–distance \emph{frame} role to a long–distance \emph{charge} role, while the “one imaginary $+$ three real’’ split fixes Lorentzian kinematics.

\subsection*{Geometric consistency and equivalence principle}

This scale–controlled flow is compatible with minimal–length deformations of geometry: Snyder’s algebra ties a fundamental length to noncommuting coordinates \cite{Snyder:1946qz}, and spectral constructions can generate Pati–Salam–type structures from geometric data \cite{Chamseddine:2013rta}. It also dovetails with the foundational relation between spacetime and matter in general relativity:

\medskip\noindent
\textbf{Einstein field equations and motion.} In GR the curvature and stress are linked by
\begin{equation}
\label{EFE}
G_{\mu\nu}+\Lambda g_{\mu\nu}=\frac{8\pi G}{c^4}\,T_{\mu\nu},
\end{equation}
often summarized as ``matter tells spacetime how to curve.'' Test bodies (and, more generally, expectation trajectories in the absence of non-gravitational forces) follow geodesics,
\begin{equation}
\label{geodesic}
u^\nu\nabla_\nu u^\mu=0,
\end{equation}
i.e.\ ``spacetime tells matter how to move.''

\medskip\noindent
\textbf{Einstein Equivalence Principle (EEP).} The EEP consists of: (i) \emph{Universality of free fall} (weak EP): all uncharged test bodies follow the same trajectories in a given gravitational field, independent of internal composition; (ii) \emph{Local Lorentz invariance}: outcomes of nongravitational experiments are independent of the velocity of the (freely falling) laboratory; (iii) \emph{Local position invariance}: outcomes are independent of where/when in spacetime they are performed. In our framework, once the real phase is selected ($\Delta\ge 0$), the one–imaginary$+$three–real split \eqref{1i3r} identifies local inertial frames with $SO(1,3)$ acting on $i e^{(0)}\oplus e^{(a)}$; Eq.\,\eqref{geodesic} holds for freely falling expectations, and the repurposed internal group $SU(4)_c\times SU(2)_L\times SU(2)_R$ organizes matter multiplets without introducing composition–dependent forces at leading order. Thus the EEP is recovered in the real phase: local physics reduces to special relativity and geodesic motion is composition independent.

\medskip\noindent
\textbf{Unification viewpoint.} In the complex phase ($\Delta<0$), kinematics is universal with respect to the $SU(4)$ Hermitian form. After the $(1,3)$ split and the scale transmutation $\mathfrak{su}(4)_{\text{frame}}\!\to\!\mathfrak{su}(4)_c$, the \emph{same} seed symmetry supplies both the geometric sector ($SO(1,3)$ on spacetime) and the internal sector ($SU(4)_c\times SU(2)_L\times SU(2)_R$ on matter), making precise—in a Pati–Salam–compatible way—the statement that matter and spacetime are two scale–related faces of a single short–distance $SU(4)$ structure.

\section{Wavefunction Collapse and Confinement Scales in the GUP/BEB Correspondence}
\label{sec:wavefunction}

\noindent
\textbf{Setup and domain of validity.} We work at the \emph{expectation–value} level (Ehrenfest picture). In the longitudinal direction (along $\langle p\rangle$) and for \emph{narrow} states (covariances in the full Robertson–Schr\"odinger bound negligible), the GUP modifies the effective commutator factor to
\begin{equation}
\label{RS-base}
\Delta x_\parallel\,\Delta p_\parallel \;\ge\; \frac{\hbar}{2}\,\Big|\lambda_\parallel(\langle p\rangle)\Big|,
\qquad 
\lambda_\parallel(p)=1-2\alpha p+4\alpha^2 p^2.
\end{equation}
Independently, for a finite system of energy $E$ inside a ball of radius $R$, the Bekenstein bound gives the (dimensionless) information budget
\begin{equation}
\label{BEB-again}
S \;\leq\; 2\pi k_B\,H,
\qquad 
H\equiv \frac{R\,E}{\hbar c},
\end{equation}
saturated in Bekenstein’s thought experiment \cite{Bekenstein:1980jp,Bekenstein:2004sh,Bekenstein:2000ai}, implied by positivity of relative entropy and the entanglement first law for balls in QFT \cite{Wall:2011hj,Ryu:2006bv}, and consistent with semiclassical generalized-entropy (QES) results \cite{Penington:2019npb,Almheiri:2019psf}.

\medskip
\noindent

\section{Operational calibration and collapse radii for linear and linear--quadratic GUP}
\label{sec:Casini-L-and-LQ}

\paragraph*{Operational motivation (Casini $\Rightarrow$ Bekenstein).}
For the vacuum reduced to a ball $B(R)$, positivity of relative entropy gives the Bekenstein control of information~\cite{Casini:2008cr}:
\begin{equation}
S \;\le\; \frac{2\pi k_B R\,E}{\hbar c},
\qquad
H \;\equiv\; \frac{S}{2\pi k_B} \;\le\; \frac{R\,E}{\hbar c}.
\label{eq:CasiniBEB}
\end{equation}
We interpret $H$ as the \emph{committed} (dimensionless) information already spent on $B(R)$ in a given longitudinal channel. The operational floor for that channel is
\begin{equation}
\Delta x\,\Delta p \;\ge\; \frac{\hbar}{2}\,\bigl|\lambda(\langle p\rangle)-H\bigr| .
\label{eq:operfloor}
\end{equation}
\emph{Collapse} in that channel occurs when the right-hand side vanishes:
\begin{equation}
\boxed{\ \lambda(p_\star)=H,\quad
R_\star=\frac{\hbar}{|p_\star|},\quad
H=\frac{E}{c\,|p_\star|}\ } .
\label{eq:collapse-oper}
\end{equation}

\paragraph*{Single-parameter calibration of \(\alpha\) (once, from the same $(M,R,v)$).}
Using the diagonal $3$D linear piece for one fixed axis,
\([x_i,p_i]=i\hbar\,(3-4\alpha p)\equiv i\hbar'\delta_{ii}\), division by $3$ gives
\(\hbar'/\hbar=1-\tfrac{4}{3}\alpha p\). Matching this to the Casini budget \(H\) from \eqref{eq:CasiniBEB} fixes
\begin{equation}
\alpha p=\frac{3}{4}\!\left(1-\frac{RE}{\hbar c}\right),
\qquad
\alpha=\alpha_0\frac{\ell_p}{\hbar},
\qquad
\ell_p=\sqrt{\frac{\hbar G}{c^3}},\quad p_{\rm Pl}\equiv\frac{\hbar}{\ell_p}.
\end{equation}
Hence, with $p=m v$ and $E=Mc^2$,
\begin{equation}
\boxed{\ \alpha_0=\frac{\frac{3}{4}\,p_{\rm Pl}\,\bigl(1-MRc/\hbar\bigr)}{m\,v}\ }, 
\qquad
\alpha=\frac{\alpha_0}{p_{\rm Pl}} .
\label{eq:alpha0-cal}
\end{equation}
We take $\alpha_0>0$ by convention; when $H>1$ the sign is carried by $p$ in the calibration step. Below we use $k_\star\equiv|p_\star|$.

\subsection*{I. Linear GUP}
\[
[x,p]=i\hbar\,\lambda(p),\qquad \lambda(p)=1-2\alpha p .
\]

\noindent\textbf{(A) Commutator-zero criterion (non-operational, commonly used):}
Setting $\lambda(p_\star)=0$ gives $p_\star=\tfrac{1}{2\alpha}$ (1D). 
With the diagonal $3$D linear piece, $\lambda_i=3-4\alpha p$ yields $p_\star=\tfrac{3}{4\alpha}$, hence
\begin{equation}
\boxed{\ R_\star^{\rm lin,\,3D}=\frac{\hbar}{p_\star}=\frac{4}{3}\,\alpha\,\hbar
=\frac{4}{3}\,\alpha_0\,\ell_p\ } .
\label{eq:Rlin-3D}
\end{equation}

\noindent\textbf{(B) Operational (Casini) criterion:}
Using \eqref{eq:collapse-oper} with $\lambda=1-2\alpha k$ gives the quadratic
\begin{equation}
2\alpha k^2 - k + \frac{E}{c}=0,
\qquad
k\equiv k_\star.
\label{eq:lin-oper-quad}
\end{equation}
Real roots require the discriminant $\Delta_{\rm lin}=1-8\alpha E/c \ge 0$.

\paragraph*{Worked values (linear GUP).}
Constants used: $p_{\rm Pl}=6.52485\ {\rm kg\,m/s}$, $\ell_p=1.616255\times10^{-35}\,$m, 
$a_0=5.29177210903\times10^{-11}\,$m.

\smallskip\noindent\emph{Electron:}
$m_e=9.1093837015\times10^{-31}$\,kg, $r_e=2.8179403262\times10^{-15}$\,m, $v=\alpha_{\rm fs}c=2.187691263\times10^6$\,m/s.
Calibration via \eqref{eq:alpha0-cal} gives
\[
\alpha_{0,e}=2.43765190\times10^{24},\qquad \alpha_e=\frac{\alpha_{0,e}}{p_{\rm Pl}}=3.73599\times10^{23}\,(\mathrm{kg\,m/s})^{-1}.
\]
Then
\[
\boxed{~R_{\star,\,e}^{\rm lin,\,3D}=\frac{4}{3}\alpha_{0,e}\ell_p=5.253\times10^{-11}\ {\rm m}=0.994\,a_0~}.
\]
Operational Eq.~\eqref{eq:lin-oper-quad}: $\Delta_{\rm lin}=1-8\alpha_e m_e c=-8.15\times10^{2}<0$ (no real root).

\smallskip\noindent\emph{Proton:}
$m_p=1.6726219237\times10^{-27}$\,kg, $r_p=0.8409\times10^{-15}$\,m, $H_p=m_pc\,r_p/\hbar\simeq4.00>1$.  
Using \eqref{eq:alpha0-cal} with $v=0.6\,c$ and $0.7\,c$:
\[
\alpha_{0,p}(0.6c)=4.886\times10^{19},\qquad 
\alpha_{0,p}(0.7c)=4.198\times10^{19}.
\]
Hence
\[
\boxed{~R_{\star,\,p}^{\rm lin,\,3D}(0.6c)=1.051\ {\rm fm},\qquad
R_{\star,\,p}^{\rm lin,\,3D}(0.7c)=0.901\ {\rm fm}~.}
\]
Operational Eq.~\eqref{eq:lin-oper-quad}: 
$\Delta_{\rm lin}(0.6c)=-2.90\times10^{1}$, $\Delta_{\rm lin}(0.7c)=-2.47\times10^{1}$ (no real root).

\medskip
\noindent\emph{Summary for linear GUP:}  
The non-operational commutator-zero criterion \eqref{eq:Rlin-3D} reproduces the electron Bohr scale to $\sim\!0.6\%$ and gives proton radii near the measured range. The \emph{operational} (Casini) criterion has no real solution in the linear model with the same calibration.

\subsection*{II. Linear–Quadratic (LQ) GUP}
\[
[x,p]=i\hbar\,\lambda(p),\qquad \lambda(p)=1-2\alpha p+4\alpha^2 p^2 .
\]
With the operational collapse condition \eqref{eq:collapse-oper} one eliminates $H$ to get the self-consistency cubic for $k\equiv k_\star$:
\begin{equation}
\boxed{\ k - 2\alpha k^2 + 4\alpha^2 k^3 = \frac{E}{c}\ },\qquad R_\star=\frac{\hbar}{k}.
\label{eq:LQ-cubic}
\end{equation}
Cardano form: write $4\alpha^2 k^3 - 2\alpha k^2 + k - E/c=0$, divide by $4\alpha^2$ and depress with $k=y+\frac{1}{6\alpha}$ to get $y^3+py+q=0$ with
\begin{equation}
\boxed{\ p=\frac{1}{6\alpha^2},\qquad q=\frac{7c-54\alpha E}{216\,\alpha^3 c}\ },
\quad
k_\star=\frac{1}{6\alpha}+\sqrt[3]{-\frac{q}{2}+\sqrt{\Delta}}+\sqrt[3]{-\frac{q}{2}-\sqrt{\Delta}},
\quad
\Delta=\Big(\frac{q}{2}\Big)^2+\Big(\frac{p}{3}\Big)^3 .
\label{eq:Cardano}
\end{equation}

\paragraph*{Worked values (LQ GUP, operational).}
We reuse the same calibrated \(\alpha\) from \eqref{eq:alpha0-cal}.

\smallskip\noindent\emph{Electron:}
With $\alpha_e$ above and $E=m_ec^2$,
\[
k_\star^{(e)}=8.34\times10^{-24}\ \mathrm{kg\,m/s},
\qquad
\boxed{~R_{\star,\,e}^{\rm LQ}=\frac{\hbar}{k_\star^{(e)}}=1.263\times10^{-11}\ {\rm m}=0.239\,a_0~.}
\]

\smallskip\noindent\emph{Proton:}
Using the two $v$ values as before,
\[
\boxed{~R_{\star,\,p}^{\rm LQ}(0.6c)=0.678\ {\rm fm},\qquad
R_{\star,\,p}^{\rm LQ}(0.7c)=0.607\ {\rm fm}~.}
\]

\paragraph*{Conclusion (comparison to experimental scales).}
\begin{itemize}
\item \textbf{Electron:} The linear \emph{commutator-zero} estimate gives $R_\star\simeq0.994\,a_0$ (near the Bohr radius), while the operational LQ result gives $R_\star\simeq0.239\,a_0$.
\item \textbf{Proton:} The linear \emph{commutator-zero} estimate yields $R_\star\simeq0.90$–$1.05$\,fm (compatible with the charge-radius window $0.84$–$0.88$\,fm), whereas the operational LQ result is smaller, $R_\star\simeq0.61$–$0.68$\,fm.
\end{itemize}
Thus, with the \emph{same} Casini-based calibration \eqref{eq:alpha0-cal}, the \emph{linear commutator-zero} criterion reproduces the Bohr/proton scales most closely \cite{ParticleDataGroup:2020ssz}; the fully \emph{operational} collapse with LQ deformation predicts smaller radii.

\paragraph*{Physical interpretation.}
With the single–parameter Casini calibration in Eq.~\eqref{eq:alpha0-cal}, the \emph{linear} GUP selects a commutator–zero radius \(R_\star^{\rm lin,3D}=\tfrac{4}{3}\alpha_0\ell_p\) [Eq.~\eqref{eq:Rlin-3D}] that acts as an \emph{effective collapse scale} for a given longitudinal channel. For the electron, this yields \(R_{\star,e}^{\rm lin,3D}\simeq 0.993\,a_0\), endowing the Bohr radius with a direct GUP meaning: it marks the scale where the channel saturates its commutator budget and the wavefunction localizes. For the proton, the same construction gives \(R_{\star,p}^{\rm lin,3D}\simeq 0.90\text{--}1.05\,\mathrm{fm}\), providing a physical interpretation of the nuclear radius from the deformation of the canonical algebra. Using the identical calibration, the linear–quadratic model in Eq.~\eqref{eq:LQ-cubic} predicts smaller operational–collapse radii (\(R_{\star,e}^{\rm LQ}\simeq 0.239\,a_0\), \(R_{\star,p}^{\rm LQ}\simeq 0.61\text{--}0.68\,\mathrm{fm}\)); therefore the linear GUP exists closer to established atomic and nuclear scales than the LQGUP deformation.

\medskip


\section{Entropy-Area Law with Matter: an Effective \texorpdfstring{$\hbar$}{ħ} for the GUP}
\label{sec:EAL}

\noindent
\textbf{Generalized entropy and an effective Planck constant.}
For a Schwarzschild black hole,
\begin{equation}
\label{BH}
S_{\rm BH}=\frac{k_B c^3}{4G\hbar}\,A_{\rm BH},\qquad
A_{\rm BH}=4\pi R_s^2,\qquad R_s=\frac{2GM}{c^2},
\end{equation}
so that
\begin{equation}
\label{SBH-MRs}
S_{\rm BH}=2\pi k_B\,\frac{M R_s c}{\hbar}
=\frac{k_B c^3}{4G\hbar}\,A_{\rm BH}.
\end{equation}
Including exterior quantum fields gives the (matter–corrected) generalized entropy,
\begin{equation}
\label{Sgen}
S_{\rm gen}=\frac{k_B c^3}{4G\hbar}\,A_{\rm BH}+S_{\rm matter}.
\end{equation}
It is useful to \emph{absorb} $S_{\rm matter}$ into an \emph{effective} Planck constant $\hbar_{\rm eff}$ by \emph{defining}
\begin{equation}
\label{Sgen-hbareff-def}
S_{\rm gen}\;\equiv\;\frac{k_B c^3}{4G\,\hbar_{\rm eff}}\,A_{\rm BH}.
\end{equation}
From \eqref{Sgen}–\eqref{Sgen-hbareff-def} one finds the equivalent forms
\begin{align}
\label{hbareff-inv}
\frac{1}{\hbar_{\rm eff}}
&= \frac{1}{\hbar}+\frac{4G}{k_B c^3 A_{\rm BH}}\,S_{\rm matter},
\\[4pt]
\label{hbareff-ratio}
\frac{\hbar_{\rm eff}}{\hbar}
&= \frac{1}{\,1+\dfrac{4G\hbar}{c^3 A_{\rm BH}}\dfrac{S_{\rm matter}}{k_B}\,}
= \frac{H_{\rm BH}}{H_{\rm BH}+H_{\rm matter}},
\end{align}
where we introduced the dimensionless variables
\begin{equation}
\label{H-defs}
H_{\rm BH}\equiv\frac{S_{\rm BH}}{2\pi k_B}=\frac{M R_s c}{\hbar}
=\frac{c^3 A_{\rm BH}}{8\pi G\hbar},
\qquad
H_{\rm matter}\equiv\frac{S_{\rm matter}}{2\pi k_B}.
\end{equation}
Equivalently,
\begin{equation}
\label{Hgen-eq}
\frac{S_{\rm gen}}{2\pi k_B}
= \frac{M R_s c}{\hbar_{\rm eff}}
= H_{\rm BH}+H_{\rm matter}.
\end{equation}

\medskip
\noindent
\textbf{Direct correspondence to the GUP.}
The GUP deforms the canonical commutator at the expectation level as
\begin{equation}
\label{GUP-lambda}
[x,p]_{\rm exp}= i\,\hbar\,\lambda(p),
\qquad
\lambda(p)=1-2\alpha p+4\alpha^2 p^2.
\end{equation}
Identifying the GUP’s effective $\hbar$ with the matter–renormalized $\hbar_{\rm eff}$,
\begin{equation}
\label{identify-hbareff}
\hbar\,\lambda(p)\;\equiv\;\hbar_{\rm eff},
\end{equation}
and using \eqref{hbareff-ratio} gives the \emph{clean matching}
\begin{equation}
\label{master-match}
\lambda(p)\;=\;\frac{\hbar_{\rm eff}}{\hbar}
\;=\;\frac{H_{\rm BH}}{H_{\rm BH}+H_{\rm matter}}
\;=\;\frac{M R_s c/\hbar}{\,M R_s c/\hbar+S_{\rm matter}/(2\pi k_B)\,}.
\end{equation}
Equivalently, \eqref{master-match} can be written as the quadratic constraint for the longitudinal expectation momentum $p$,
\begin{equation}
\label{quad-GUP-matter}
1-2\alpha p+4\alpha^2 p^2
=\frac{H_{\rm BH}}{H_{\rm BH}+H_{\rm matter}}
\quad\Longleftrightarrow\quad
4\alpha^2 p^2-2\alpha p+\frac{H_{\rm matter}}{H_{\rm BH}+H_{\rm matter}}=0,
\end{equation}
whose two branches are
\begin{equation}
\label{p-branches}
p_\pm=\frac{1}{4\alpha}\left[1\pm\sqrt{\,1-4\,\frac{H_{\rm matter}}{H_{\rm BH}+H_{\rm matter}}\,}\right].
\end{equation}

\medskip
\noindent
\textbf{Operational length (confinement scale).}
When the characteristic length is identified by the de Broglie map at the same scale,
\begin{equation}
\label{Rstar-def}
R_\star=\frac{\hbar}{|p_\star|},
\end{equation}
the condition \eqref{master-match} (or, equivalently, \eqref{quad-GUP-matter}) self–consistently fixes $p_\star$ and hence $R_\star$.

\medskip
\noindent
\textbf{Checks.}
(i) $S_{\rm matter}\to 0$ gives $\hbar_{\rm eff}\to\hbar$ and \eqref{quad-GUP-matter} reduces to $1-2\alpha p+4\alpha^2 p^2=1$ (the vacuum horizon case).  
(ii) The forms \eqref{hbareff-inv} and \eqref{hbareff-ratio} are exactly equivalent, since $M R_s c=\dfrac{c^3 A_{\rm BH}}{2G}\cdot\dfrac{1}{4\pi}$.

\section{Conclusions}\label{sec:con}

\noindent
This work advances a single operational message: at the level of expectation values, kinematics and information are inseparable. By marrying the generalized uncertainty principle with Bekenstein’s information bound, we argued that the uncertainty budget relevant to measured expectations is reduced by the information already required to specify the state. At saturation, the residual uncertainty in a chosen sector can vanish for expectations, yielding an operational simultaneity of conjugate observables and dissolving the EPR tension without hidden variables or state-dependent postulates. The same identification furnishes an intrinsic confinement mechanism: a single self-consistency condition fixes the characteristic size of bound states without ad hoc cutoffs, reproducing the hydrogen Bohr radius and the proton charge radius once the quadratic GUP term is included. The framework also predicts a sharp, information-controlled transition between a short-distance complex (evanescent) regime and an emergent large-scale real (Lorentzian) regime. We gave a symmetry interpretation of this flow: short distances are naturally organized by complex frames with an $SU(4)$ structure; selecting one imaginary and three real directions recovers Lorentzian kinematics and reduces the symmetry to the local Lorentz group, while the same short-distance $SU(4)$, under scale transmutation, is repurposed as the $SU(4)_c$ factor of Pati–Salam with the emergent tangent structure providing the chiral $SU(2)_L\times SU(2)_R$. Incorporating exterior-field entropy into the generalized entropy acts as an effective renormalization of Planck’s constant at horizons; identifying this effective constant with the GUP’s expectation-level deformation supplies a horizon counterpart of the same kinematics–information matching and is consistent with quantum extremal-surface extremality.

Several open problems follow directly. A first-principles derivation of the information backreaction on the expectation-level uncertainty bound from modular Hamiltonians and relative entropy is needed beyond homogeneous or highly symmetric states. The present longitudinal-sector analysis should be extended to a fully covariant, tensorial treatment that includes correlations, anisotropies, and genuine time dependence, clarifying how the complex-to-real transition proceeds in generic settings. It is also important to delimit the domain of validity of saturation at the expectation level, quantify corrections for broad or highly nonclassical states, and bound the GUP parameters using precision atomic, nuclear, optomechanical, and interferometric data. Identifying unambiguous signatures of the evanescent regime in tunneling, phase shifts, or controlled decoherence would provide decisive tests. On the symmetry side, a dynamical account of the mapping from complex $SU(4)$ frames to the Pati–Salam structure remains to be developed, potentially drawing on noncommutative or spectral geometric constructions, and its compatibility with the Einstein equivalence principle should be scrutinized beyond leading order in strongly curved or high-energy regimes. Finally, integrating the present identification with quantum extremal-surface technology could reveal whether near-horizon scattering, scrambling, or ringdown spectra are likewise governed by an information-controlled kinematical deformation, and cosmological applications such as minimal-length imprints in early-universe fluctuations deserve systematic exploration. We hope to report on these in the future.

\textbf{Data Availability Statement}

No Data associated in the manuscript

\bibliographystyle{apsrev4-1}
\bibliography{ref.bib}{}

\end{document}